\documentclass[a4paper,twocolumn,showpacs,superscriptaddress,floatfix,nofootinbib]{revtex4-1}
\renewcommand{\nat}{{Nature}}
\renewcommand{\apj}{{ApJ}}
\renewcommand{\prd}{{PRD}}

\newcommand{\apjl}{{ApJL}}
\newcommand{\cqg}{{CQG}}
\newcommand{\mnras}{{MNRAS}}

\usepackage{latexsym}
\usepackage{amssymb}
\usepackage{amsfonts}
\usepackage{amsmath}
\usepackage{bm}
\usepackage{graphicx}   
\usepackage{color}
\usepackage{subfigure}
\usepackage{times}
\usepackage{units}
\usepackage{hyperref}
\usepackage{filecontents}

\begin{document}

\title{Magnetically-induced outflows from binary neutron star merger remnants}

\author{Daniel M. Siegel}
\affiliation{Max Planck Institute for Gravitational Physics (Albert Einstein Institute), Am M\"uhlenberg 1, 14476 Potsdam-Golm, Germany}

\author{Riccardo Ciolfi}
\affiliation{Physics Department, University of Trento, Via Sommarive 14, 38123 Trento, Italy} 
\affiliation{INFN-TIFPA, Trento Institute for Fundamental Physics and Applications, Via Sommarive 14, 38123 Trento, Italy}

\begin{abstract}        
Recent observations by the \textit{Swift} satellite have
revealed long-lasting ($\sim 10^2-10^5\,\mathrm{s}$), ``plateau-like'' X-ray afterglows in the vast
majority of short gamma-ray bursts events. This has put forward the
idea of a long-lived millisecond magnetar central engine being generated in a
binary neutron star (BNS) merger and being responsible for the sustained
energy injection over these timescales (``magnetar model''). We elaborate here on
recent simulations that investigate the early evolution of such a
merger remnant in general-relativistic
magnetohydrodynamics. These simulations reveal very different conditions than
those usually assumed for dipole spin-down emission in the magnetar
model. In particular, the surrounding of the
newly formed NS is polluted by baryons due to a dense, highly
magnetized and isotropic wind from the stellar surface that is induced by magnetic field
amplification in the interior of the star. The timescales and
luminosities of this wind are compatible with early X-ray afterglows,
such as the ``extended emission''. These isotropic winds are a generic
feature of BNS merger remnants and thus represent an attractive
alternative to current models of early X-ray afterglows. Further
implications to BNS mergers and short gamma-ray bursts are discussed.
\end{abstract}        

\maketitle

\section{Introduction}\label{sec:introduction}

\noindent Coalescing binary neutron stars (BNSs) and neutron star-black hole
(NS-BH) binaries are not only among the most promising candidates for
direct detection of gravitational waves (GWs) with upcoming
ground-based interferometers such as advanced LIGO and Virgo
\cite{Harry2010}. The accretion of a remnant torus around
a BH formed promptly after merger also represents the leading scenario
to explain the formation of a short gamma-ray burst (SGRB; e.g.,
\cite{Eichler1989}). However,
the recent discovery of long-lasting X-ray
afterglows (referred to as ``extended emission'' and ``X-ray
plateaus''; e.g., \cite{Rowlinson2013}) in the vast
majority of SGRB events observed by the
\textit{Swift} satellite \cite{Gehrels2004} indicate ongoing energy
injection on timescales of up to $\sim 10^{4}\,\mathrm{s}$, which are
incompatible with the short accretion timescale of a remnant torus
($\lesssim 1\,\mathrm{s}$).

These X-ray afterglows can, however, be interpreted by assuming that
the merger leads to the formation of a long-lived, rapidly rotating NS
(a millisecond magnetar; henceforth referred to as the ``magnetar
model''; e.g., \cite{Metzger2008}). Despite
individual differences in the phenomenology, different variants of the
magnetar model have in common that they postulate the formation of a
\textit{uniformly} rotating, highly magnetized NS and that at least the late-time X-ray emission
is powered by dipole spin-down radiation, which requires an essentially baryon-free environment around the NS. 

The magnetar model challenges the NS-BH progenitor channel for SGRBs,
since a magnetar cannot be formed in this case, and we are thus led
to consider BNS mergers as the main formation channel. The outcome of
a BNS merger depends on the equation of state of nuclear matter at
high densities, which is unknown. However, the recent discovery of
high-mass neutron stars \cite{Demorest2010}
indicate a maximum mass for stable NSs of $M_\mathrm{TOV}\gtrsim 2\,
\mathrm{M}_\odot$. Furthermore, the mass distribution of BNSs is peaked around
$1.3-1.4\,\mathrm{M}_\odot$ and this yields an approximate mass of the merger
remnant of $\approx 2.4\,\mathrm{M}_\odot$ \cite{Belczynski2008}. Combining
this with the fact that uniform rotation can centrifugally support NSs
against gravitational collapse up to masses of $\approx
1.2\,M_\mathrm{TOV}\gtrsim 2.4\,\mathrm{M}_\odot$ \cite{Lasota1996}, merger
remnants are very likely supramassive. If the merger remnant is only
slightly hypermassive (i.e. above the maximum mass for uniformly
rotating NSs), it can still migrate to a supramassive star by mass
loss through early winds while still supported by differential
rotation (see below). Supramassive NSs are typically
long-lived and the exact lifetime depends on the timescale to remove a
significant fraction of the rotational energy (e.g., via dipole
spin down). Hence, it is conceivable that in the vast majority of
BNS mergers a long-lived NS is formed, which supports the magnetar model.

From simulations it is known, however, that at early times the BNS
merger remnant is characterized by a high
degree of differential rotation and thus substantial magnetic field
rearrangement and amplification. These
conditions are very different from pure dipole spin down of a
uniformly rotating magnetar, as assumed in most magnetar
models. This leads us to investigate this early phase of a
differentially rotating BNS merger remnant by fully
general-relativistic magnetohydrodynamic simulations. We elaborate
here on recent results of such simulations \cite{Siegel2014}, which
revealed the generation of a very dense, highly magnetized wind that
carries sufficient Poynting flux to explain typical luminosities of early
X-ray afterglows. The present paper focuses on discussing the wind
properties in more detail and its implications for SGRBs and their
X-ray afterglows.

\begin{figure*}[t]
\centering 
\includegraphics[angle=0,width=0.95\textwidth]{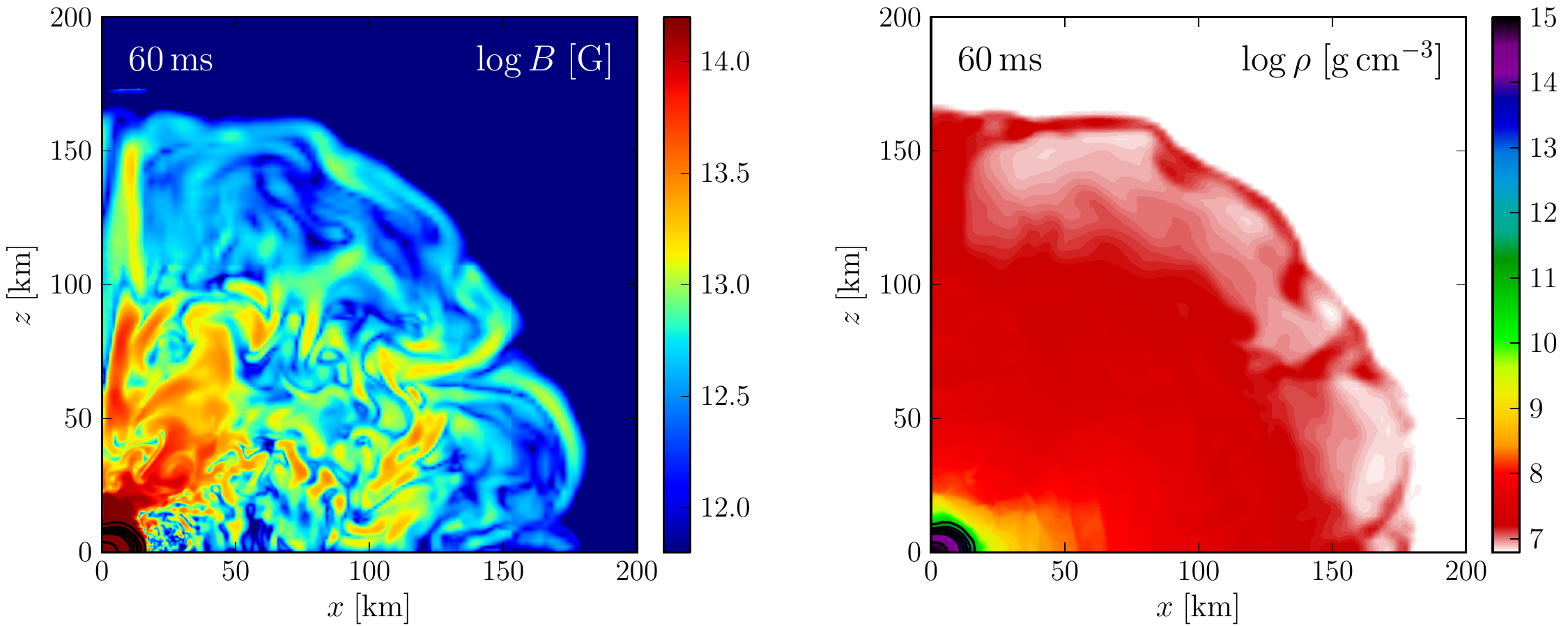}
\caption{Magnetic field strength (left) and rest-mass density (right)
  of the wind from a differentially rotating NS (corresponding to model \texttt{rand}
  of \cite{Siegel2014}; the star is indicated
  by black density contours in the lower-left corners). These snapshots
  showing a very dense, highly isotropic and highly magnetized wind are
  taken at the end of the simulation.}
\label{fig:wind2D} 
\end{figure*}

\section{Initial data and numerical setup}

\noindent Employing the general-relativistic ideal magnetohydrodynamics code \textsc{WhiskyMHD}
\cite{Giacomazzo2007} combined with the
publicly available Einstein Toolkit and its spacetime evolution code
\textsc{McLachlan} \cite{Loeffler2012}, models of
hypermassive neutron stars constructed with the \textsc{RNS} code
\cite{Stergioulas1995} were evolved, which we endowed with different initial
magnetic field geometries (for details, see
\cite{Siegel2014}). These models are representative of the typical
outcome of a BNS merger, once the merger remnant has settled down to a
roughly axisymmetric state and assuming that the merger has not led to
a prompt collapse to a BH. Here, we
elaborate, in particular, on model \texttt{rand} of \cite{Siegel2014},
which has a ``random'' initial magnetic field geometry inside and
outside the star and which we consider
to most closely resemble the actual magnetic field geometry resulting
from a BNS merger. It is important to stress that our
conclusions in this paper and that of \cite{Siegel2014} mostly depend
on differential rotation and thus also apply to differentially
rotating supramassive NSs (which together with hypermassive NS are the
most likely result of BNS mergers, see Section~\ref{sec:introduction}).

\section{Evolution, mass ejection, electromagnetic emission}

\noindent The early evolution of a differentially rotating BNS merger remnant is
characterized by magnetic winding: poloidal magnetic fields in the
interior of the star are ``wound up'', producing a strong toroidal
component and enhancing the magnetic pressure. Rotational energy
is thus converted into magnetic energy. We find that
within a few average rotational periods, the increase of magnetic pressure in the
surface layers of the star is sufficient to overcome the
gravitational binding of material in the vicinity of the stellar
surface. This results in a highly magnetized, highly isotropic and
very dense wind of matter from the NS that heavily pollutes the environment with baryons
(cf. Figure~\ref{fig:wind2D}).\footnote{In the
case of a dominant dipole moment in the initial magnetic field
structure, we expect an additional collimated wind component directed
along the rotation axis \cite{Siegel2014}.} This environment of the NS
is very different from an essentially
baryon-fee pulsar wind that is usually hypothesized in magnetar models.

\begin{figure*}[t]
\centering 
\includegraphics[angle=0,width=\textwidth]{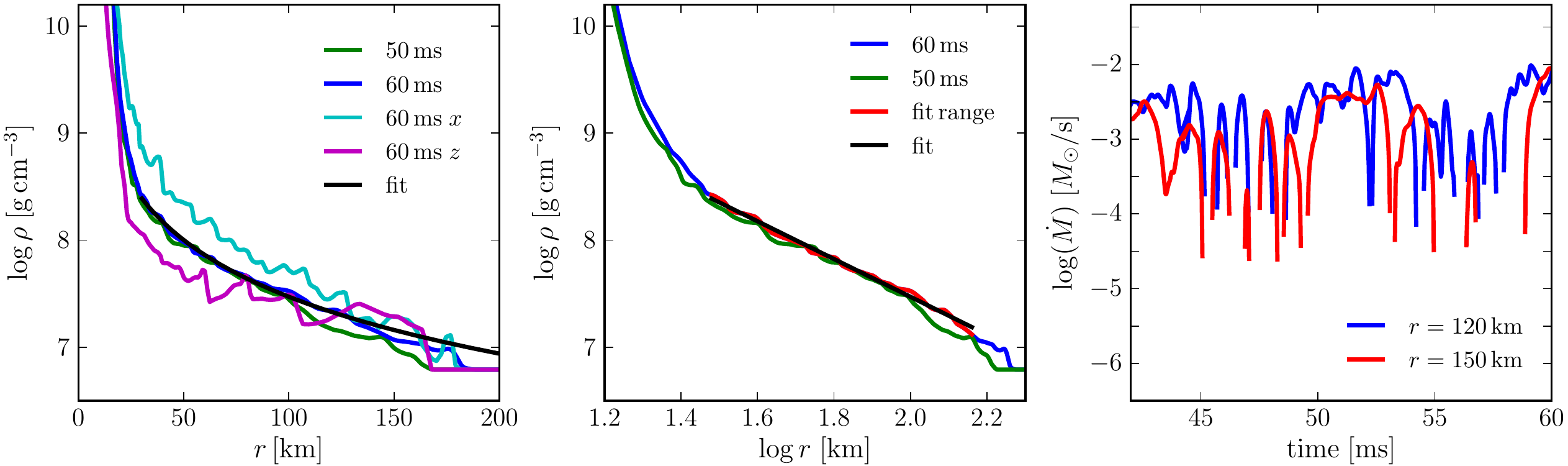}
\caption{Left and middle panel: rest-mass density profiles of the wind from a differentially
  rotating NS (corresponding to model \texttt{rand} of
  \cite{Siegel2014}) at different times towards the end of the
  evolution (density profiles along the $x$ and $z$ axes in cyan and
  purple, respectively, polar angle averaged profiles in green and
  blue). The black
line shows a fit $\rho\propto r^{-2}$ to the angular averaged
profiles. Right panel: instantaneous mass-loss rate associated to the wind as
measured at different radii during the evolution. }
\label{fig:rho_vel} 
\end{figure*}

A global measure of the baryon pollution is depicted in the rightmost panel of Figure~\ref{fig:rho_vel}, which shows the total mass crossing a coordinate
sphere of radius $r=120\,\mathrm{km}$ and $r=150\,\mathrm{km}$ from the
centre of the star per unit
time during the evolution of the simulation. The wind at these
distances is still rather turbulent during the later part of the
evolution, which results in a noisy mass-loss rate. Nevertheless, the
mass-loss rates as measured at different
distances from the star roughly agree and yield a stationary mass-loss
rate of a few in $10^{-3}$ solar masses per second ($\dot{M}\sim
10^{-3}\,\mathrm{M}_\odot\,\mathrm{s}^{-1}$).

The morphology of the wind is depicted in the left and middle panel of
Figure~\ref{fig:rho_vel} in terms of the rest-mass density
profiles. The directionally averaged profiles\footnote{The system is
  very close to axisymmetry and averaging over the polar angle is
  sufficient to obtain a directionally averaged profile.} at different times
(e.g., at
50\,ms and 60\,ms) agree very well up to the outer radius of the bulk of the ejected matter
at that particular time. A fit to these profiles
reveals a characteristic average wind density profile
$\rho\propto r^{-2}$ that is constant in time. According to mass
conservation, we have at fixed radius
\begin{equation}
  \dot{M}\equiv \left|\frac{\partial M(r,t)}{\partial t} \right|= 4\pi r^2\rho(r,t) v(r,t),
\end{equation}
where $M$ is the amount of mass inside the volume of radius $r$ and
$v(r,t)$ is the radial velocity of the wind at time $t$ and radius
$r$. Combining $\rho(r,t) = \rho(r)\propto r^{-2}$ and $\dot{M}(r,t) =
\dot{M}=\mathrm{const}$ thus yields $v(r,t)=v_0=\mathrm{const}$,
i.e. an overall bulk speed of the wind that is constant in time and
space. Assuming $\dot{M}\simeq 5\times 10^{-3}\,\mathrm{M}_\odot$ (see above
and Figure~\ref{fig:rho_vel}), we obtain a subrelativistic wind
expansion speed for the model considered, $v_0\simeq 0.03\,c$, where $c$ is the speed of light.

Although the mass-loss rate is roughly constant during the timescale
of up to 60\,ms that we can access through our simulations, it is expected
to change significantly at later times. The timescale for a change of
$\dot{M}$ and thus for a change of the wind properties depends on the
timescale for removal of differential rotation (see
Section~\ref{sec:consequences}). Furthermore, neutrinos radiated from
the hot interior of the NS can heat and drive an outflow from the
surface of the NS and thus lead to additional mass loss (e.g.,
\cite{Metzger2008a}). Depending on the magnetic field
strength and the rotation period, however, the magnetically-driven
wind discussed here may well be the dominant process for
mass loss.

The Poynting flux associated to the magnetically-driven wind
in our simulations corresponds to stationary electromagnetic (EM) luminosities
\begin{equation}
  L_\mathrm{EM}\simeq 10^{48}\,\bar{B}_{15}^{2}\,R_\mathrm{e,6}^3\,
  P_{-4}^{-1}\,\mathrm{erg}\,\mathrm{s}^{-1}, \label{eq:luminosity}
\end{equation}
where $\bar{B}$ is the average magnetic field strength in the outer
layers of the star once the system has reached its stationary state,
$R_\mathrm{e}$ denotes the equatorial radius, and $P$ is the (central)
spin period \cite{Siegel2014}. This relation is universal in the sense that it does not
depend on the initial magnetic field geometry and it is different
from dipole spin-down emission considered in the magnetar model, in
which case $L_\mathrm{EM}\propto B_\mathrm{p}^2\,R_\mathrm{e}^6\, P^{-4}$,
where $B_\mathrm{p}$ is the dipolar field strength at the pole. Assuming
that only a small fraction $\eta \equiv
L_\mathrm{EM}^\mathrm{obs}/L_\mathrm{EM}\sim 0.01-0.1$ of the luminosity
(Equation~\ref{eq:luminosity}) is converted into X-rays, the observed X-ray
luminosities $L_\mathrm{EM}^\mathrm{obs}\sim
10^{46}-10^{51}\,\mathrm{erg}\,\mathrm{s}^{-1}$ of extended emission
and X-ray plateaus of SGRBs (e.g., \cite{Rowlinson2013})
are obtained by magnetic field strengths $\bar{B}\sim
10^{14}-10^{17}\,\mathrm{G}$, which are likely produced during and
after a BNS merger \cite{Siegel2014,Siegel2013}.

\section{Conclusions: Consequences for SGRBs and X-ray afterglows}\label{sec:consequences}

\noindent The timescale over which the baryon-loaded wind and its associated EM
luminosity can be sustained depends on the timescale for removal of
differential rotation and is uncertain. An order of magnitude estimate
is provided by the Alfv\'en timescale \cite{Shapiro2000}, which corresponds to $\sim 1-10\,\mathrm{s}$  for the magnetic fields
considered here and which is in
agreement with an estimate $\Omega/\dot{\Omega}\lesssim 10\,\mathrm{s}$
for the timescale of change of the
angular velocity profile in our simulations. Differential rotation is completely
removed within a few to several of these timescales.
The overall duration of the process as well as the luminosities mentioned above
are compatible with the observed extended emission of SGRBs (lasting
$\lesssim 100\,\mathrm{s}$) and thus make the
magnetically-driven wind a promising alternative model to explain such
emission.

The prompt SGRB emission is thought to be collimated and thus beamed
away from us in the vast majority of events. Furthermore, the observed SGRB
rate within the sensitivity volume of advanced LIGO and Virgo is very
small ($\lesssim 1\,\mathrm{yr}^{-1}$ \cite{Metzger2012}). Therefore,
identifying potential isotropic EM emission from BNS mergers is of utmost
importance. The isotropic EM emission found here potentially
represents such an important counterpart to the GW
signal of BNS mergers.

Although the initial phase of differential rotation of the BNS merger
remnant is rather short, the total amount of mass lost via the
magnetically-driven wind can be substantial. The ejected matter
creates an optically thick environment that will likely affect the
emission properties of the system in the following evolution, once the
star has settled down to uniform rotation. At this stage, the NS might lose
rotational energy via spin-down radiation, and the resulting X-ray
afterglow radiation would be the product of a reprocessing of this radiation
by the confining envelope of ejecta matter
(cf. \cite{Ciolfi2015}). The associated spectrum could
carry such a signature, e.g., thermalization of (part of) the non-thermal spin-down emission by the
optically thick ejecta material.

As the presence of a magnetically-driven wind only depends on
differential rotation, which is generic to BNS merger remnants, the associated
baryon pollution in the polar regions is also generally expected. This
severely hampers the production of a relativistic jet through accretion of a remnant torus around the newly
formed NS. Hence, if the presence of long-lasting X-ray
afterglows requires a long-lived NS remnant, an alternative explanation
for the generation of the SGRB itself is likely required. Such a
scenario that explains the production of long-lasting X-ray afterglows and the
prompt SGRB emission in a common phenomenology has recently been proposed
\cite{Ciolfi2015} (for an alternative proposal, see \cite{RezzollaKumar}). However, full simulations of BNS
mergers will be needed in the future to better characterize the magnetically-driven
wind. These simulations will also
help to shed light on whether or not a relativistic jet can be
launched at the merger of a BNS system forming a long-lived NS.


\end{document}